\documentclass{article}
\usepackage{adjustbox}
\usepackage{amsmath}
\usepackage{amsfonts}
\usepackage{amssymb}
\usepackage[utf8]{inputenc}
\usepackage{graphicx}
\usepackage{color}
\usepackage{url}
\usepackage{array}
\usepackage{subfigure}

\newcommand\Tstrut{\rule{0pt}{2.5ex}}         % = `top' strut
\newcommand{\footremember}[2]{%
    \footnote{#2}
    \newcounter{#1}
    \setcounter{#1}{\value{footnote}}%
}
\newcommand{\footrecall}[1]{%
    \footnotemark[\value{#1}]%
}
\def\R{{\mathbb R}}

\newcommand{\pool}[1]{\underset{#1}{pooling}}

\title{Steganalysis via a Convolutional Neural Network using Large Convolution Filters \\
for Embedding Process with Same Stego Key \\
\textit{Erratum note\footnote{
This is a corrected version of the eprint 
\textit{J.-F. Couchot, R. Couturier, C. Guyeux, and M. Salomon (2016): Steganalysis via a Convolutional Neural Network using Large Convolution Filters. Publication: eprint arXiv:1605.07946. Publication Date: 05/2016}. In this version, we used improperly prepared stego images with the same stego key, resulting in embedding changes pretty much in the same places. This error has been pointed out by a reviewer after submission to IEEE Transactions on Information Forensics \& Security.
}}
}

\author{
\begin{tabular}[t]{c@{\extracolsep{2.5em}}c}
Jean-Fran\c{c}ois Couchot\footremember{alley}{FEMTO-ST Institute, UMR 6174 CNRS - University Bourgogne Franche-Comt\'e, Belfort, France.
DISC (department of computer science and complex systems) - AND team} & Rapha\"el Couturier\footrecall{alley} \footnote{raphael.couturier@univ-fcomte.fr}\\
Christophe Guyeux\footrecall{alley} & Michel Salomon\footrecall{alley} \Tstrut
\end{tabular}
}

\date{\today}

\begin{document}

\maketitle

\begin{abstract}

For the past few years, in the race between image steganography and steganalysis, deep learning has emerged as a very promising alternative to steganalyzer approaches based on rich image models combined with ensemble classifiers. A key knowledge of image steganalyzer, which combines relevant image features and innovative classification procedures, can be deduced by a deep learning approach called Convolutional Neural Networks (CNN). These kind of deep learning networks is so well-suited for classification tasks based on the detection of variations in 2D shapes that it is the state-of-the-art in many image recognition problems. In this article, we design a CNN-based steganalyzer for images obtained by applying steganography with a unique embedding key. This one is quite different from the previous study of {\em Qian et al.} and its successor, namely {\em Pibre et al.} The proposed architecture embeds less convolutions, with much larger filters in the final convolutional layer, and is more general: it is able to deal with larger images and lower payloads. For the ``same embedding key" scenario, our proposal outperforms all other steganalyzers, in particular the existing CNN-based ones, and defeats many state-of-the-art image steganography schemes.

\vfill

\end{abstract}

\section{Introduction}

The aim of steganography is to embed a private message into an innocent cover inside a public communication channel and to finally extract the message at the communication target. A steganographic scheme is thus composed of two algorithms: the embedding one and the extraction one. The security of a steganographic scheme is evaluated according to its ability to be invisible. The less the cover is modified to embed the hidden message, the more secure the approach is. The slightly modified cover that contains the message is usually referred to as stego content. Highly Undetectable steGO (HUGO)~\cite{DBLP:conf/ih/PevnyFB10}, WOW~\cite{conf/wifs/HolubF12}, UNIWARD~\cite{HFD14}, STABYLO~\cite{DBLP:journals/adt/CouchotCG15}, EAI-LSBM~\cite{5411758}, and MVG~\cite{FK2013} are some of the  most efficient instances of such schemes when the cover is an image. In all the previously aforementioned schemes, a distortion function returns a large value in a smooth area which can be easily modeled and a small one in textured, ``chaotic'' area, \textit{i.e.}, where there is no obvious model.

Steganalysis aims at discovering whether an object contains (or not) a hidden message. This work focuses on image steganalysis, \textit{i.e.}, when cover objects are images. It takes place into the context where the image domain is known. Additional knowledge on the embedding algorithms and the payload can help the steganalyzer tool. Practically speaking, most steganalyzers are twofold machine learning tools. The objective of the first step is to capture as much information as possible, on images it has to deal with, and through the computation of a set of features. Many image features have been published: Rich Models (RM) have been proposed for the spatial domain (SRM)~\cite{DBLP:journals/tifs/FridrichK12,DBLP:journals/tifs/HolubF13,DFH14} or for the JPEG one~\cite{HF15,HF15c}. The second step is a machine learning approach that aims at distinguishing stego images from cover ones according to specific values of selected features. To the best of our knowledge, tools that are based on Ensemble Classifier~\cite{DBLP:journals/tifs/KodovskyFH12}, further denoted as EC, are considered as a state of the art steganalyzer. This one implements a fusing step (decision function) based on a simple majority voting between Fisher Linear Discriminant (FLD) components. The two keys of EC are the FLD and the features it embeds. A Support Vector Machine \cite{schaathun2012svm} or a Multilayer Perceptron (MLP)\cite{sabeti2010mlp,Lubenko:2012:SMC:2361407.2361410}~can~serve~as~a~suitable~alternative.

Today, for many classification tasks the state-of-the-art methods belong to deep learning \cite{SIG-039,Schmidhuber201585,lecun2015deep}, a kind of machine learning approach that has been receiving a continuously increasing attention in the past decade. Deep learning encompasses approaches that try to automatically extract the most relevant high-level features of the input data, to improve the learning of the targeted task~\cite{6472238}. Typically, a deep learning method consists in a network~/~ set of connected processing layers, where in each layer multiple transformations (mainly nonlinear) are performed. The different layers are usually organized in a feedforward manner, but recurrent networks with feedback connections are also possible. Among the various deep learning approaches we can notice Deep Belief Networks \cite{hinton2006fast}, Convolutional Neural Networks (CNN or ConvNet) \cite{yoo2015deep}, and even, to some extent, the so-called Reservoir Computing~\cite{verstraeten2007experimental,lukovsevivcius2009reservoir} (RC, like Echo State Networks \cite{jaeger2001short,Lukosevicius2012} and Liquid State Machines \cite{maass2002real,maass2010liquid}). Furthermore, this vision of Reservoir Computing networks as a part of the Deep Learning paradigm makes sense with the emergence of deep Reservoir Computing Networks (RCN) obtained by stacking multiple reservoirs. For example, in \cite{jalalvand2015real} Jalalvand {\em et al.} applied to the MNIST image classification problem a deep RCN of three reservoirs, each of them composed of 16K neurons, leading to 528K trainable parameters.

Let us notice that, in the case of the Reservoir Computing paradigm, various hardware implementations have already been realized. This is due to the much simpler learning operation that consists in a linear regression scheme, even if a deep RCN is a deep Recurrent Neural Network (RNN) \cite{hermans2013training}. In particular, several optical reservoir computers have achieved, with good performances, various tasks like spoken digit recognition, image classification, or channel equalization. Such a neuromorphic computing platform has been proposed more recently for convolutional networks by IBM using a TrueNorth chip \cite{2016arXiv160308270E}, a new chip architecture based on spiking neurons. The main interest of neuromorphic computing is the energy-efficiency and the high throughput of such implementation, even if, sometimes, the problem-solving performance is slightly lower than a state-of-the-art computer simulation. For example, Larger {\em et al.} \cite{Larger:12} have proposed an optoelectronic reservoir computer able to recognize 1,000,000 spoken digits per second almost without any error (Word Error Rate of $0.04\pm0.017$\%), while IBM's chip can classify images and speech sequences at a throughput between 1,100 and 2,300 frames per second. The high-speed photonic implementation proposed by Brunner {\em et al.} \cite{brunner2013parallel} further improved the performances obtained in \cite{Larger:12}: a lower error (WER of $0.014\%$) at a simultaneously highest data rate (1.1 GByte/s), showing that in comparison with an hybrid optical-electronic system, an all-optical neuromorphic processor can offer higher throughput. 

Deep learning architectures are particularly fruitful for solving classification or recognition problems on datasets of images or videos. Therefore, since the purpose of a steganalyzer is to perform image classification (detect if a message is embedded or not in it), we focus our attention on  convolutional neural networks \cite{lecun1998gradient} based deep learning decision procedures. Such deep learning networks are the state-of-the-art for many classical standard datasets, like MNIST \cite{wan2013regularization}, CIFAR-10 \cite{2014arXiv1412.6071G}, or CIFAR-100~\cite{2015arXiv151107289C}. In comparison, for the MNIST problem the deep RCN proposed in \cite{jalalvand2015real} has achieved an error rate of 0.92 percent, whereas the state-of-the-art method \cite{wan2013regularization} (a CNN with a dropout approach in its fully connected part performing the classification task) reaches an error rate more than four times lower. Deep ConvNets are not only impressive with toy datasets like MNIST, they are also very successful when dealing with real life data such as images provided by medical imaging technologies. In this latter field of application, deep learning appears to be a breakthrough technology for fast and automated early detection of illnesses. Identifying signs of diabetic retinopathy in eye images, which is a major cause of blindness and vision loss, is a good illustration of the great positive impact of deep learning \cite{DreamUpV}, since it allows a real-time detection of eye disorders with an accuracy similar to that of ophtalmologists. Note that the Kaggle Diabetic Retinopathy Detection competition (2015) was won by a SparseConvNet proposed by Ben Graham similar to the state-of-the-art ones he proposed for the CIFAR tasks \cite{2014arXiv1412.6071G,2014arXiv1409.6070G}.

In fact, a CNN-based steganalyzer would allow to automatically unify feature extraction and classification steps in one unique architecture, without any a priori feature selection. The first experiments of a CNN architecture, using stacked convolutional auto-encoders done by Tan and Li~\cite{tan2014stacked}, have not reached a level of accuracy similar to the one given by SRM+EC-based steganalyzers. Further works~\cite{qian2015deep} succeeded to have comparable performance when applied on images of size $256\times 256$ with large payloads: from 0.3 to 0.5 bit per pixel (bpp). Finally,
this previous work has been recently improved by Pibre {\em et al.} \cite{pibre2016deep} in the context of the scenario where stego images are always obtained by using the same embedding key, leading to a slightly different and less complex CNN (with fewer layers, but larger ones) able to divide by three the detection errors for a 0.4~bpp payload. In this work we also consider the ``same embedding key" scenario but, as noticed by Pibre {\em et al.}, let us emphasize that this scenario is not recommended because embedding several messages with the same key weakens security, since in that case embedding changes occur rather in the same locations. Even if these successive works have shown the relevance of CNN-based steganalysis, some limitations have still to be overcome: processing larger images, in other domains (\textit{e.g.}, in JPEG one), and with other payload values (smaller ones)%, such as 0.05 bpp for example).  \textcolor{red}{Raph: cet exemple est dangereux, non? on ne l'a pas testé}

The objective of this article is to design a CNN-based steganalyzer that further improves the performance of the previous works. Moreover, our proposal aims at being more general and at overcoming the limitations noted previously. More precisely, the contributions of this work can be summarized as follows:
\begin{itemize}
\item Firstly, even if the proposal consists of a CNN as in the previous research works \cite{qian2015deep} and \cite{pibre2016deep}, the proposed network is quite different from those ones. On the one hand we have less convolutional layers and on the other hand the final fully connected part doing the classification task is reduced to its simplest form.
\item Secondly, the CNN we introduce is more general, able to process larger images, to detect steganography tools that embed messages in the spatial \textit{and} the frequency domain, and with lower payload values.
\end{itemize}
As will be shown thereafter, a key idea is the use of large convolution filters (almost as large as the image to process) to build features giving a high level abstraction of the input data. Thanks to this principle our deep learning network gives very good results. 

The remainder of this research work is organized as follows. In Section~\ref{sec:otherworks} we review existing works related to steganalyzer design. We start by giving some details on Ensemble Classifier-based approaches and we outline the CNN-based ones. The next section is devoted, after a brief overview of the convolutional neural networks architecture, to the description of the proposed CNN architecture. Section~\ref{sec:expe} shows the relevance of the proposed convolutional network through various experimental scenarios. The considered datatsets and parameter settings are first described and then the results obtained with various scenarios and payloads are introduced. In Section~\ref{sec:discuss}, we discuss the benefits of the proposed approach and the points that need to be further investigated. This article ends by a conclusion section, in which the contributions are summarized and intended future work is outlined.

\section{Related works} \label{sec:otherworks}

This section presents the state-of-the-art results in steganalysis. It first recalls the most accurate results with conventional approaches~(see Section~\ref{sub:art:classical}), which is followed by deep learning based steganalysis results~(see Section~\ref{sub:art:deep}). 

\subsection{Conventional steganalysis}\label{sub:art:classical}

Let us first focus on steganalysis when the embedding scheme works in a spatial domain. 
Ensemble classifier~\cite{DBLP:journals/tifs/KodovskyFH12} and Rich Models~\cite{doi:10.1117/12.872279} have been formerly combined in~\cite{DBLP:journals/tifs/FridrichK12}. This combination allows to detect the steganographic algorithm called HUGO~\cite{DBLP:conf/ih/PevnyFB10} with a
detection error of 0.13 (resp. of 0.37) with a payload of 0.4 (resp. 0.1) bit per pixel
on the BOSS base. Such rich models are based on co-occurrences of noise residuals between a possibly modified pixel and its neighborhood ones. However, such an approach gives exponential increasing of the co-occurrence matrix size with respect to the neighborhood length. This issue has been addressed in~\cite{DBLP:journals/tifs/HolubF13} where the authors enlarge neighboring residual sizes, and project them into random directions instead of extracting co-occurrence matrices. This projection allows to reduce the number of errors: images possibly modified by the HUGO scheme are detected with an error of 0.12 (resp of 0.36) with a payload of 0.4 (resp. 0.1) bit per pixel on the BOSS database. For the same set of images, the detection errors of the WOW~\cite{conf/wifs/HolubF12} steganographic scheme is 0.18 (resp. 0.39) with a payload of 0.4 (resp. 0.1)~bit per pixel. For the S-UNIWARD~\cite{HFD14} scheme, the errors are respectively 0.18 and 0.40 for these payloads. The authors of~\cite{Tang:2014:ASA:2600918.2600935} have proposed to select a subset of suspicious pixels according to the known payload. The SRM features are further extracted and the decision is achieved by Ensemble Classifier. Thanks to this supplementary knowledge, the detection errors for the WOW scheme are reduced to 0.18 and to 0.34 for payloads of 0.4 and 0.1 bpp respectively. To the best of our knowledge, the most accurate results in spatial domain steganalysis have been obtained in~\cite{DBLP:conf/wifs/DenemarkSHCF14}. In their work, the authors modify the calculus of the co-occurrence matrix by memorizing the maximum of the neighboring change probabilities instead of their mean (but still across 4~residuals). With this modification, leading to a much larger scan direction, the average steganalysis errors for the WOW scheme are reduced to 0.15 and to 0.30 for payloads of 0.4 and 0.1~bpp respectively. Moreover, the S-UNIWARD~\cite{HFD14} scheme errors are  0.19 and 0.37 respectively for these payloads.

Projection of residuals on random directions~\cite{DBLP:journals/tifs/HolubF13} in SRM has been applied too in the JPEG domain and is denoted as JPSRM features. With this set of features, the authors obtained detection errors of 0.43 and 0.13 for payloads of 0.1 and 0.4~bpp respectively, with J-UNIWARD and a JPEG Quality Factor set to 0.75.
DCTR features~\cite{DBLP:journals/tifs/HolubF15} have been introduced as an alternative 
to projection in random directions. In the features calculus, projections are achieved with 
DCT bases. This change of bases allows to reduce the size of the feature set  
while preserving its accuracy. When applied on J-UNIWARD with 0.75 JPEG Quality Factor,
detection errors are 0.44 and 0.13 for payloads of 0.1 and 0.4~bpp respectively.
To the best of our knowledge, the most trustworthy feature set is PHARM~\cite{HF15c}. It is a continuation of the DCTR features in the sense that it still takes the position of the residual into the DCT grid. Compared to this previous set, PHARM only considers small prediction kernels to compute residuals. With 0.75 JPEG Quality Factor, detection errors are 0.42 and 0.12 for payloads 0.1 and 0.45 with PHARM features.

\subsection{Deep learning based steganalysis}\label{sub:art:deep}

The design of steganalyzers based on a deep learning network, a powerful machine learning technique that has become a breakthrough technology as noted in the previous section, has recently been investigated. More precisely, the use of a convolutional neural network (a deep learning approach matching exactly the underlying two-step process in classical steganalysers) to fulfill the steganalysis task, has outperformed conventional methods like the ones described in the previous paragraph. Thereafter, we will present works that have investigated the use of such networks to detect if an image embeds a secret message when using various steganographic algorithms, like HUGO~\cite{DBLP:conf/ih/PevnyFB10}, WOW~\cite{conf/wifs/HolubF12}, and S-UNIWARD~\cite{HFD14} in \cite{qian2015deep}, or only the last one in \cite{pibre2016deep}. For the experiments, both works considered databases of $256 \times 256$~pixels grey-level images, mainly from BOSS database, but also in~\cite{pibre2016deep} using an homemade database called LIRMMBase, which has been obtained by mixing Columbia, Dresden, Photex, and Raise databases.

First of all, in~2015 Qian {\em et al.} \cite{qian2015deep} proposed a CNN consisting of 5~convolutional layers finally producing 256~features followed by three fully connected layers (for the classification part): two~hidden layers of 128~ReLU (for Rectified Linear Unit) neurons each and 2~softmax~neurons in the output layer (a more detailed overview of CNNs architecture is given in the next section). Besides, this CNN does not process an input image directly, but rather works on a $252 \times 252$~high-pass filtered image issued by a $5\times 5$ kernel denoted $F^{(0)}$. The experiments showed that the designed CNN was only slightly outperformed by state-of-the-art SRM+EC-based steganalyzers. In fact, even if they obtained a lower detection accuracy (a few percent larger errors: 3\% to 4\%), they emphasized that a CNN is a promising way for steganalysis. 

Later, Pibre {\em et al.} \cite{pibre2016deep} investigated this former work further and improved the detection performance when embedding key is reused for different images. They obtained a CNN with a different shape: fewer but larger layers, able to reduce the detection error by more than 16\% in comparison with the state-of-the-art in case of embeddings with S-UNIWARD at 0.4~bpp. Hence, the features are extracted by 2 layers with 64~convolution kernels in the first layer and 16~kernels in the second one, using no subsampling and ReLU neurons, while in \cite{qian2015deep} they had 5~layers of 16~kernels each and Gaussian neurons. As a consequence, the number of features increased dramatically from 256 to 258,064~features, even with slightly larger kernels. The authors claim that, compared to the dramatic reduction of the input image in 256~features in \cite{qian2015deep}, using less numerous but larger convolutional layers is one of the reasons why they obtained better results. Obviously, the huge number of features to process by the fully connected part also needs larger hidden layers, therefore they used 1,000~ReLU neurons in both of them. Let us also emphasize the fact that for Pibre {\em et al.} the filtering step with $F^{(0)}$ is mandatory for steganalysis: without it, they observed that none of the CNNs they tested was able to converge. They also evaluated the classification ability of a neural network restricted to the fully connected part, in other words a classical MLP with two hidden layers (2,000~ReLU neurons in each one), considering as input the high-pass filtered version of an image. As they obtained results which were only slightly worse than with a CNN (clairvoyant scenario detection accuracy: 24.67\% for RM+EC, 7.4\% with a CNN, and 8.75\% for the feedforward MLP), the benefit of the convolutional layers appears not to be minor in comparison with the preliminary high-pass filtering. Unfortunatly, when considering a different key for each embedding they obtained bad results. In fact, in that case the detection error reaches 45.31\% meaning that the CNN is not able to identify a pattern specific to stego images. We will mostly compare our approach with this work and the conclusions drawn by it, since we considered the same scenario than them where a single embedding key is used. Like Pibre {\em et al.}, and may be Qian {\em et al.}, this choice results from the use of the C++ implementations of steganographic algorithms for the embedding which can be downloaded from DDE Lab Binghamton web site.

\section{Principle of the method} \label{sec:method}

\subsection{Convolutional neural network architecture overview}

A convolutional neural network consists of one or several convolutional layers, followed by some fully connected layers of neurons like in classical multilayer feedforward neural networks. The background idea of the convolutional layers is to learn how to extract sets of smaller feature maps with kernels from 2D input data, so that the final maps, processed by the fully connected part doing the classification, give a better representation of the original data. On the one hand it reduces the input's dimensionality and on the other hand it automatically finds the most suitable features to allow the CNN to fulfill the expected task. That explains the success of convolutional neural networks in many image or video recognition problems and why they are the current state-of-the-art in image classification tasks.

\begin{figure}[t] 
\centering
\begin{tabular}{cc}
(a) & \adjustimage{width=0.9\linewidth,valign=m}{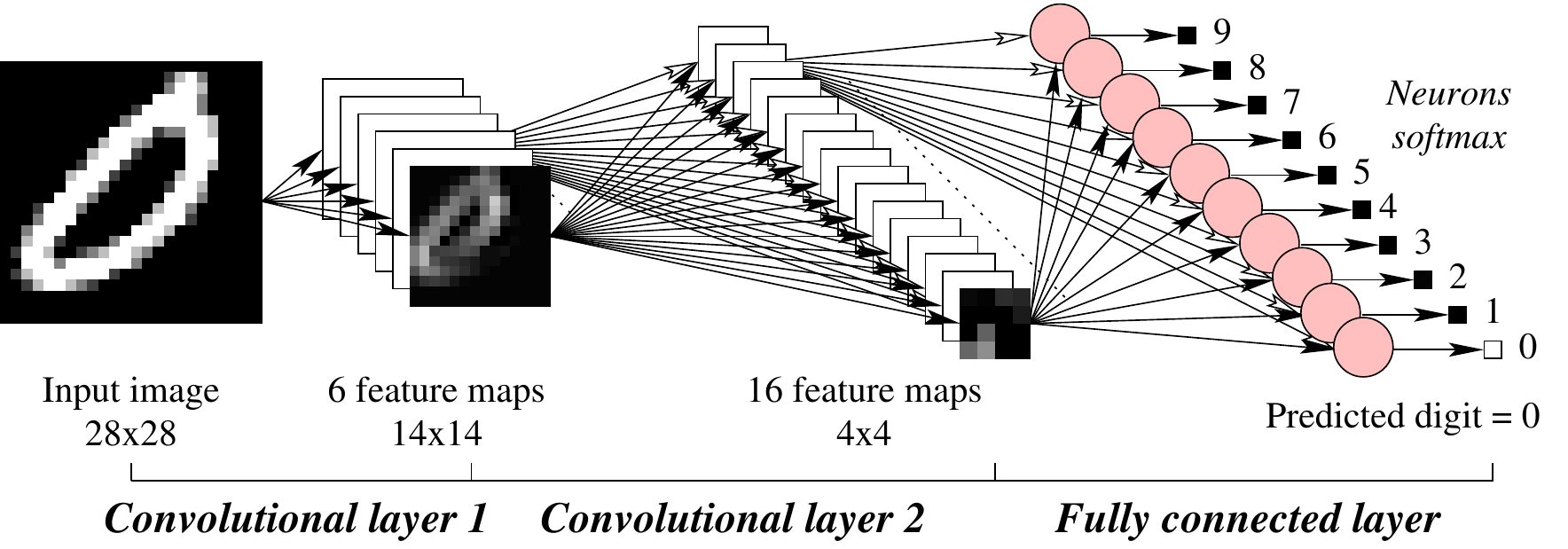} \\
\smallskip \\
(b) & \adjustimage{width=0.9\linewidth,valign=m}{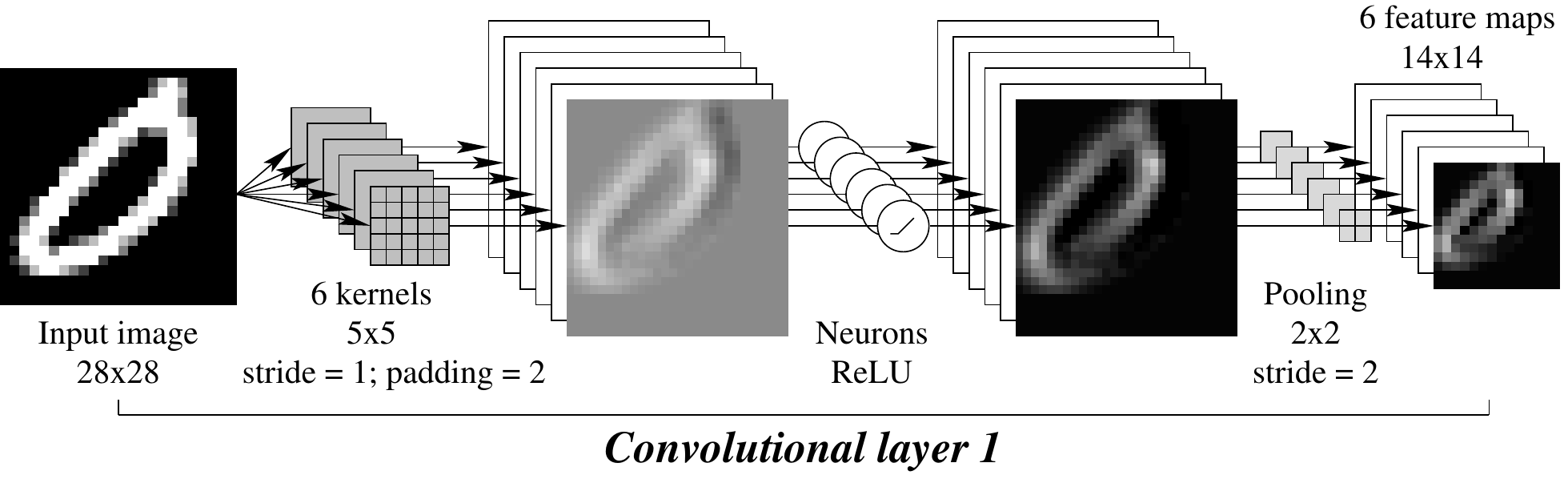}
\end{tabular}
\caption{A convolutional neural network for the MNIST problem: global architecture (a) and detailed view of the first convolutional layer (b).}
\label{fig:mnist}
\end{figure}

Each convolutional layer usually produces feature maps by a three-step process. The first step performs some filtering using $K$ kernels leading to $K$ new feature maps. Therefore each kernel is applied on the existing feature maps resulting from the previous convolutional layer, or on the initial two dimensional input data in the case of the first convolutional layer, and the new feature map is the linear combination of the filtered maps. Then, another step might reduce each feature map thanks to a pooling operation, typically by computing the mean or the max over $p \times p$ regions. % (with a value of 2 for $p$).
A final step adds some nonlinearity with the introduction of a layer of $K$ neurons before or after the subsampling step. In comparison with fully connected layers, convolutional layers have a lower training cost, since the weights to optimize during the learning process to obtain a feature map are the corresponding kernel values and neuron biases. The sharing of the kernel values by all the data points not only reduces the set of weights, but also improves the generalization performance. As an illustration, Figure~\ref{fig:mnist}(a) shows a CNN architecture for the MNIST classification problem. The MNIST is a large database of $28 \times 28$ pixels grey-level images obtained by normalizing and centering handwritten digits scans. As can be seen, the CNN consists of two convolutional layers, producing respectively 6 and 16~feature maps, and a fully connected part reduced to the output layer with one output neuron per possible digit value.

Similarly in other neural networks, the training (or learning) process consists in minimizing a training error (or loss function) using an optimization algorithm that updates the network parameters (weights and biases). A batch gradient-based optimization algorithm is the most common choice and the Stochastic Gradient Descent (SGD) is very popular, particularly in the deep learning community, but more advanced techniques, such as AdaDelta or AdaGrad, might also be considered. Obviously, in that case the well-known backpropagation algorithm, which allows to compute the gradient of the training error, is the workhorse of the training process. The network parameters are usually optimized until the number of updates exceeds some specified maximum values, or when the training error reaches a given limit. 

Let us now give some further details on the different steps of the proposal.% applied in a convolutional layer.

\begin{itemize}
\item[$\bullet$] Convolution

If we denote by $C^{kl}$ the result in layer~$l$ of the convolution with the $k$-th kernel defined by the weighting matrix $W^{kl}$, then we have:
\begin{equation}
C^{kl} = \sum_{m=1}^{K^{l-1}} \left( W^{kl} \ast F^{m(l-1)} \right),
\end{equation}
where $\ast$ denotes the usual convolution product, $K^{l-1}$ is the number of kernels in the previous layer, and $F^{m(l-1)}$ the $m$-th final feature map produced by the latter. For the first convolutional layer, {\it i.e.}, when $l=1$, $K^{l-1}=K^0=1$, and $F^{1(l-1)}=F^{10}=I$ is the input image. The size of the matrix filter $W^{kl}$ implicitly defines  the size of the local area, also denoted as \emph{receptive field} in analogy with the visual system, used to compute the value $C^{kl}(i,j)$ with coordinate $(i,j)$ in $C^{kl}$.

The size of the output of the convolution also depends on two parameters called stride and padding.
The first one controls the overlapping between two neighboring receptive fields in each spatial dimension (width and height).
A stride value $S$ means that the convolution operation is applied by spatially shifting the weighing matrix $W^{kl}$ from $S$ units. The second parameter allows to pad the borders of the input feature map~$F^{m(l-1)}$ or image~$I$ with zeros. Thus, let be given a stride value $S$, and a padding value $P$, we obtain for a square input image~/~map $F^{m(l-1)}$ an output $C^{kl}$ such that:
\begin{equation}
dim\left(C^{kl}\right)=\left( dim\left(F^{m(l-1)}\right) - dim\left(W^{kl}\right)+2 \times P\right)/S+1.
\end{equation}
In particular, to keep for $C^{kl}$ the same size as the input data $F^{m(l-1)}$ ($dim\left(C^{kl}\right)=dim\left(F^{m(l-1)}\right)$), the stride and padding values must be set to $1$ and $\lfloor dim(W^{kl})/2 \rfloor$, respectively.

\item[$\bullet$] Activation function

In order to add some nonlinearity, each $C^{kl}$ is processed by an activation function $f^{kl}: \R \rightarrow \R$ similar to those used in other neural networks, considering usually a similar function $f^{kl}$ for the $K^l$ filters. Various choices are possible for $f^{kl}$, such as classical sigmoid functions like the logistic function or the hyperbolic tangent $f(x)=\tanh(x)$, 
a Gaussian function defined by: 
\begin{equation}\label{eq:Gaussian-func}
f(x)= \dfrac{e^{-x^2}}{\sigma^{2}},
\end{equation}
or the ReLU (Rectified Linear Unit) function defined by:
\begin{equation}\label{eq:RELU}
f(x)=\max(0,x),
\end{equation}
\noindent and so on. Among them, the ReLU is a notable choice for the convolutional layers in CNN, this one was for example chosen by Pibre {\em et al.} \cite{pibre2016deep} whereas Qian {\em et al.} \cite{qian2015deep} have chosen a less conventional Gaussian function (see Equation~(\ref{eq:Gaussian-func})). When the training process is gradient-based, the activation function must be differentiable. For each kernel a bias value~$b^{kl}$ also allows to control the shifting of the activation function, therefore we have $K^l$ additional parameters to optimize in layer~$l$. To sum up, after this step a map $C^{kl}$ is transformed into $A^{kl}$ as follows:
\begin{equation}
A^{kl} = f^{kl} \left(C^{kl}+b^{kl}\right).
\end{equation}

\item[$\bullet$] Pooling (or subsampling)

During this step, a 2D map is down-sampled by being partitioned it into regions of $p_l~\times~p_l$ values, where each region is replaced by the mean or the maximum of its values. A stride value might allow to control the overlapping between neighboring regions, even if the choice of non-overlapping ones is widespread (stride value of $p_l$). The value of $p_l$ is usually picked in the range of $2$ to $5$, according to the size of the input data of layer~$l$. The background idea of the pooling is twofold: reducing the variance across the values and reducing the size of the feature map. 

In the context of steganalysis, the choice of the average operation in \cite{qian2015deep} is motivated by the low stego noise. However, despite experiments showing that in comparison with max pooling the average operation is the best choice, in \cite{pibre2016deep} they decided to drop the pooling step  considering that it leads  loss of information. This choice partly explains the huge  number of features in output of the convolutional part in the CNN proposed in \cite{pibre2016deep} and its relevance is assessed by their experiments.

\end{itemize}

Figure~\ref{fig:mnist}(b) precisely shows  the chaining of the different steps in the case of the CNN considered for the MNIST problem, by giving a detailed view of its first convolutional layer. It can be seen that $K^1=6$~convolution kernels of size $5 \times 5$ are applied with stride and padding values equal, respectively, to $1$ and $2$, and thus produce 6~maps having a similar size to that of the input image. Each of these maps is then processed by a ReLU neuron (see Equation (\ref{eq:RELU})), followed by a max pooling with non-overlapping regions of $2 \times 2$~values (stride with value~$2$), finally leading to a feature map $F^{k1}$ that has a size of one quarter of the input image. Formally, each final feature map $F^{k1}$ in output of the first convolutional layer satisfies, for $k=1,\dots,K^1=6$:
\begin{eqnarray}
F^{k1} & = & \pool{max}\left(A^{k1}\right) = \pool{max}\left(f^{k1} \left(C^{k1}+b^{k1}\right)\right) \nonumber \\
\end{eqnarray}

\begin{eqnarray}
F^{k1} & = & \pool{max}\left(f^{k1} \left(\sum_{m=1}^{K^0} \left( W^{k1} \ast F^{m(l-1)}        \right)+b^{k1}\right)\right) \\
 & = & \pool{max}\left(f^{k1} \left(W^{k1} \ast F^{10}+b^{k1}\right)\right) = \pool{max}\left(f^{k1} \left(W^{k1} \ast I+b^{k1}\right)\right) \nonumber
\end{eqnarray}
where $I$ is the original input image.

\subsection{Shape of the proposed convolutional neural network}

The design of the proposed convolutional neural network, described in details thereafter, was driven by the following considerations. 

Firstly, in \cite{qian2015deep} and \cite{pibre2016deep} there is no indisputable proof of the optimality of the kernel $F^{(0)}$ used to first filter the input image, and which more or less operates as an edge detection filter. In fact, in \cite{pibre2016deep} the authors only experimentally observed that, without the high-pass filter~$F^{(0)}$, the CNNs they studied were not able to converge, and thus they considered it as a prerequesite to application of CNNs to steganalysis. Therefore, a question is why could his kernel not be learned by the CNN? 

Secondly, steganographic algorithms embed the secret message by modifying pixels that are correlated and widespread throughout the whole input image. Consequently, we think that it is better to use large convolution filters to build features able to highlight the slight underlying modifications performed by a steganographic algorithm. Various filter sizes can be found in the literature, going from most common sizes $3 \times 3$ and $5 \times 5$, to $12 \times 12$ or $15 \times 15$. For example, in the case of the MNIST problem, which deals with images of $28 \times 28$~pixels, the filters on the first convolution layer usually have a size of $5 \times 5$. Larger filters are more suited for images containing more complex information like natural images. Overall, the choice of the filters depends on the input dataset and the expected data correlations which will guide the classification process.

%In order to make our network we followed a different approach from paper [ref chinois] and [ref lirmm]. Both of these research articles consider a small convolution filter called $F^{(0)}$ that more or less operates as an edge detection phase. Then some small convolutional filters are used. 

%As steganalizers work with an entire image and not only small parts, we think that it is better to use big convolutional filters. Hence, they should be able to detect any kind of steganalyzer. We choose to replace the  constant filter (defined by hand) $F^{(0)}$ by a small convolutional filter and at the end we use a classical softmax layer that is used to decide if the image is a cover or a stego. 

%So compared to the other works we use only three layers and our architecture is quite different from the other approaches.

%Various filter sizes can be found in the literature, going from most common sizes 3x3 and 5x5, to 12x12 or 15x15. For example, in the case of the MNIST problem, which deals with images of 28x28 pixels, the filters on the first layer has usually a shape of 5x5. Larger filters are more suited for images containing more complex information like natural images. In fact, the objective of the filters in the convolution layers is to build successively relevant abstractions of the images which allow the final fully connected MLP to perform the classification task. Obviously, the choice of the filters depends on the dataset and the expected data correlations which will guide the classification process. 

\begin{figure}[t!] %[htbp!]
\centering
\includegraphics[width=\linewidth]{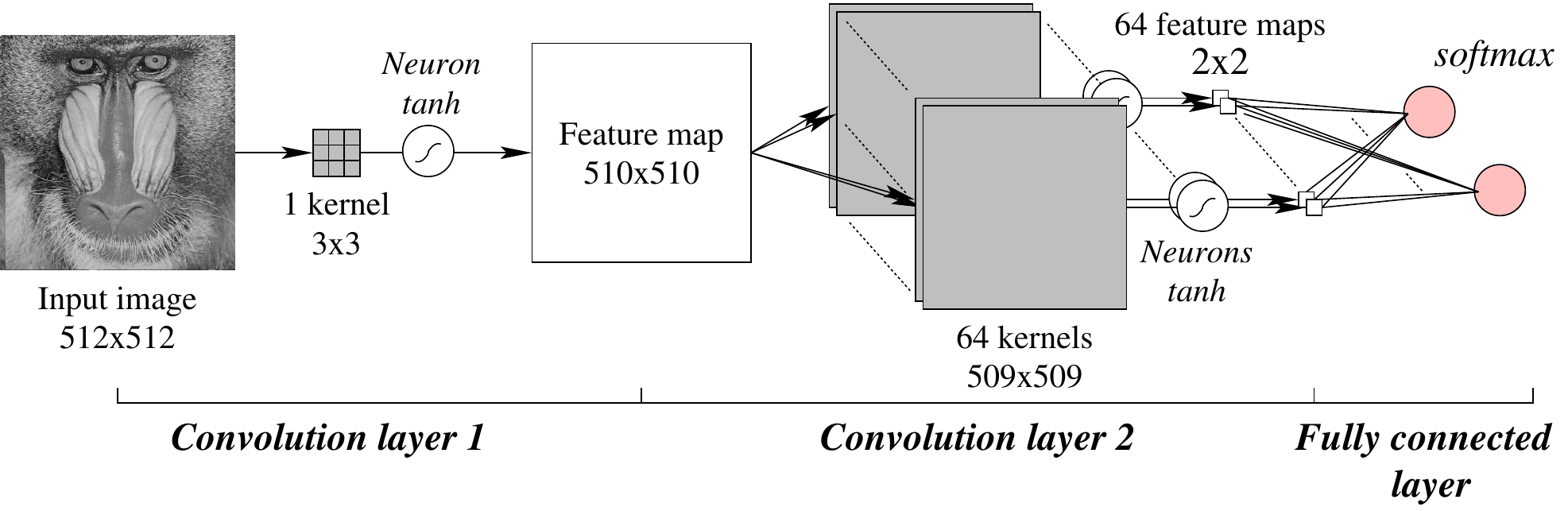}
\caption{Convolutional neural network proposed for steganalysis.}
\label{fig:filter}
\end{figure}

Practically, with the previous guidelines in mind and after some preliminary experiments, we retained an architecture quite different from the ones proposed in \cite{qian2015deep} and \cite{pibre2016deep}, as detailed in Figure~\ref{fig:filter}. The convolution part consists of two layers with hyperbolic tangent function as activation function. The first one reduced to a single kernel of size $3~\times~3$ to achieve a first filtering in a way similar to $F^{(0)}$, followed by a layer of 64~filters as large as possible with zero-padding (a stride of~1). As we consider $512 \times 512$~pixels input images, the filtered image $F^{11}$ issued by layer~1 is a $510 \times 510$ image and the 64~final feature maps $F^{k2}$, $1 \leq k \leq 64$, given by layer~2 are of size $2 \times 2$, since the filters are such that $\dim\left(W^{k2}\right)=509$. Compared to \cite{qian2015deep} the convolutional part of the proposed CNN results in the same number of features ($256$~features), but with less convolutional layers as in \cite{pibre2016deep} and an input image twice larger in both axes. Note that the pooling operation is dropped in both layers. The fully connected part is a classical neural network in its  simplest form: a single output layer of two~softmax~neurons. This is a major difference with the CNN designed in the previous works \cite{qian2015deep,pibre2016deep}. The reason why this minimal fully connected network with no hidden layer is able to fulfill the classification task and detect successively images with a hidden message, as shown in the next section, is the relevance of the proposed convolution part architecture for steganalysis.

Using torch~\cite{collobert2011torch7} as deep learning platform, the model corresponding to the CNN described above is
simply defined by:
\begin{verbatim}
model:add(nn.SpatialConvolutionMM(1, 1, 3, 3))
model:add(nn.Tanh())
model:add(nn.SpatialConvolutionMM(1, 64, 509, 509))
model:add(nn.Tanh())
model:add(nn.Reshape(64*4))
model:add(nn.Linear(64*4, 2))
model:add(nn.LogSoftMax())
\end{verbatim}
The source code used to carry out the experimental assessment in the next section, with some pretrained networks, is available for download from {\tt GitHub}\footnote{\url{https://github.com/rcouturier/steganalysis_with_deep_learning.git}}.

%The complete source code used to carry out the experimental assessment in the next section, with some pretrained networks, is available for download from {\tt GitHub} in the following repository:
%\begin{center}
%\url{https://github.com/rcouturier/steganalysis_with_deep_learning.git}
%\end{center}
%Let us note that the interested reader can easily access to pretrained CNNs in the {\tt GitHub} repository, in order to be able to reproduce the experiments presented thereafter.

\section{Experimental results} \label{sec:expe}

%In this section, we evaluate the detection performance of the proposed convolutional neural network. We start
In order to evaluate the detection performance of the proposed convolutional neural network, we need cover images, embedding schemes, and to set several parameters. Thus, we start this section by a description of the input image datasets, the selected embedding schemes (steganographic algorithms), and the parameters setup, in particular the parameters relative to the training method. Then we present the steganalysis detection performance for two scenarios similar to the ones considered in \cite{pibre2016deep}. As performance metric, we chose the detection (or classification) accuracy: the percentage of correctly classified image samples, which will be computed on a set of images different from the ones used to train the CNN (disjoint testing and training sets). In practice, the detection accuracy is measured by means of the confusion matrix provided by the Torch tool.

\subsection{Datasets and parameters setting}

The experiments were performed considering images provided by two image cover databases. The first database is the well-known BOSS one \cite{conf/ih/PevnyFB10} ({\it Break Our Steganographic System}), which consists of $512~\times~512$~grayscale images, while the second one is the Raise database \cite{conf/mmsys/Dang-NguyenPCB15}. These databases are standard ones, built to evaluate detection algorithms. Currently only three steganographic tools have been tested and thus used to produce stego images: WOW~\cite{conf/wifs/HolubF12}, HUGO~\cite{DBLP:conf/ih/PevnyFB10}, and J-UNIWARD~\cite{HFD14}. Both former ones work in the spatial domain, whereas the latter works in the frequency (JPEG) domain. Moreover, the different steganographic algorithms are applied to generate stego images with two embedding payload values: 0.4 and 0.1~bpp, using the C++ implementations available from DDE Lab Binghamton web site\footnote{\url{http://dde.binghamton.edu/download/stego_algorithms/}}.

%\textcolor{blue}{Michel:Je pense que ce serait mieux de mettre ça dans la partie discussion, juste avant la remarque du Raphus (je l'ai déjà fait)?}
% First of all,  Table~\ref{tab:payloads} summarizes the average number of pixels that have been modified with respect to selected payload, on the 10,000 BOSS $512\times 512$ images. 
% For instance, when the payload is set to $\alpha=0.1$ with HUGO, about only 2,425 pixels are modified in each image whereas 
% $0.1 \times 512\times 512$ is 26,214. The difference is mainly due to the use of the Syndrome Treillis Code~\cite{DBLP:conf/mediaforensics/FillerJF10}. 

% \begin{table}
%     \centering
%     \begin{tabular}{|l|l|l|l|}
%         \hline
%          payload & WOW & HUGO & J-UNIWARD \\
%         \hline
%          0.1     &  2,349    &  2,425    &  727 \\
%         \hline
%          0.4     & 11,730    &  11,778    & 3,591         \\
%         \hline
%     \end{tabular}
%     \caption{Average number of modified bits w.r.t payloads}
%     \label{tab:payloads}
% \end{table}

Compared to~\cite{pibre2016deep}, we made more exhaustive evaluation experiments, since Pibre~{\em et al.} considered a single steganographic algorithm: S-UNIWARD, using only the larger payload value to hide information in the stego images. It should also be noticed that the pixel values of images are first divided by $255$, and then globally normalized using the mean and the standard deviation computed across all the images.

We have provided a detailed description of the proposed CNN architecture in the previous section, the only point that needs to be further discussed is the tuning of the training (or learning) parameters. Several optimization algorithms are available to train a neural network among which gradient-based ones are the most common choice. In this work, we chose to apply a mini-batch stochastic gradient descent (SGD), a very popular learning method in deep learning. This algorithm updates the network parameters after the processing of a mini-batch of training samples, using the gradient values computed thanks to the backpropagation algorithm. A mini-batch gradient descent is interesting because it usually leads to a more stable convergence. In the experiments, the training parameters were set as follows: a mini-batch size of $100$~samples, a learning rate value of $0.5$ with a $5 \times 10^{-7}$ decay, and a zero momentum.

To control the duration of the training process, various criteria can be considered, independently or together. For example, the training might be performed for a fixed number of network parameters updates (or epochs), until a predefined training error is reached, and so on. In this work, we retained as stopping criteria a maximum number of training epochs. In fact, the training should continue as long as the network generalization ability is improved and the overfitting avoided. According to Hinton, an early stopping approach, like the one described in \cite{prechelt1998early}, is an appealing approach to monitor the training process.

From a computational complexity point of view, the cost of the training depends on numerous parameters such as the number of input data samples to be processed during an epoch, the number of epochs, the number of executed operations required to obtain the network output for one input sample, and so on. However, a key parameter is the number of network parameters (weights and biases) to be optimized. In the case of the proposed CNN this number is huge due to the $64$~filters of size $509 \times 509$ to be optimized in the second convolutional layer: $64 \times (509 \times 509+1)=16,581,248$~parameters. In comparison, the first convolutional layer and the fully connected part induce a tiny number of parameters: $1 \times (3 \times 3+1)$~parameters for layer one and $2 \times (64+1)$ ~parameters associated to the two output neurons. Fortunately, the computational cost resulting from a such huge number of parameters to optimize is not as prohibitive as it used to be the past, thanks to the various computing platform available today. GPU platforms like the NVIDIA Tesla~K40~GPU used in this work are  particularly interesting, as they are optimized to execute the kind of operations used in CNNs and more generally in deep learning.

\subsection{Scenarios}

\begin{itemize}
\item Scenario~1

This scenario can be seen as a version of the clairvoyant scenario considered in \cite{pibre2016deep}. On the one hand, the cover and stego images come from the same database and on the other hand we know which steganographic algorithm is used, just as the payload value with which the hidden message is embedded. Thus, for the training of the network only the first 4,000~images of the BOSSBase \cite{conf/ih/PevnyFB10} are used, whereas for the testing stage 2,000 images have been used. %(images numbered 7,001 to 9,000).
Recall that three different steganographic algorithms are used: WOW~\cite{conf/wifs/HolubF12}, HUGO~\cite{DBLP:conf/ih/PevnyFB10}, and J-UNIWARD~\cite{HFD14}, with, in each case, two payload values.

\begin{figure}%
\centering
\subfigure[][WOW, $\alpha=0.4$~bpp]{%
\label{fig:accuracy_WOW_0.4}%
%\label{fig:ex3-a}%
\includegraphics[scale=0.39]{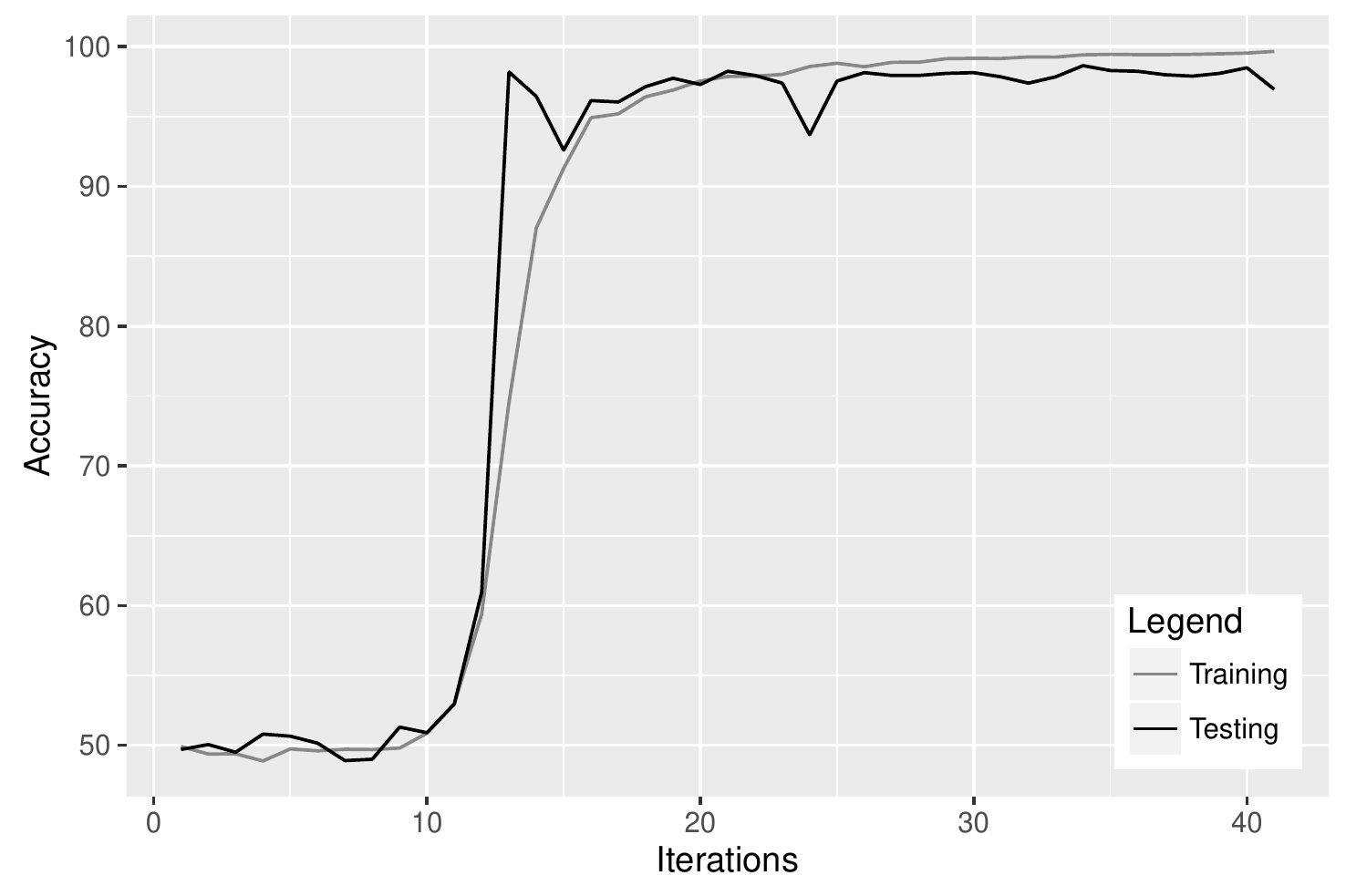}
}%
\hspace{4pt}%
\subfigure[][WOW, $\alpha=0.1$~bpp]{%
\label{fig:accuracy_WOW_0.1}%
\includegraphics[scale=0.39]{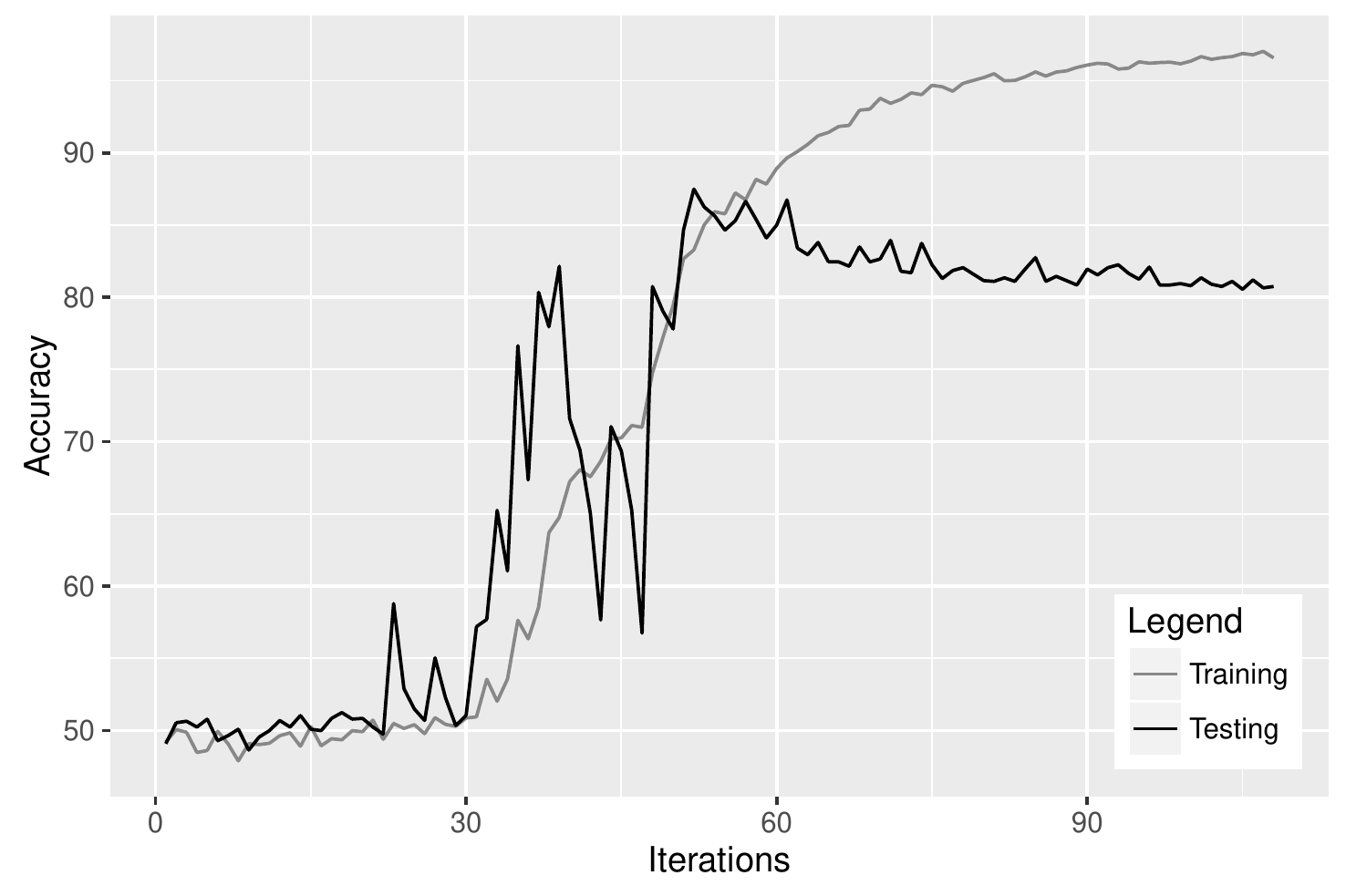}
}
\\
\subfigure[][HUGO, $\alpha=0.4$~bpp]{%
\label{fig:accuracy_HUGO_0.4}%
\includegraphics[scale=0.39]{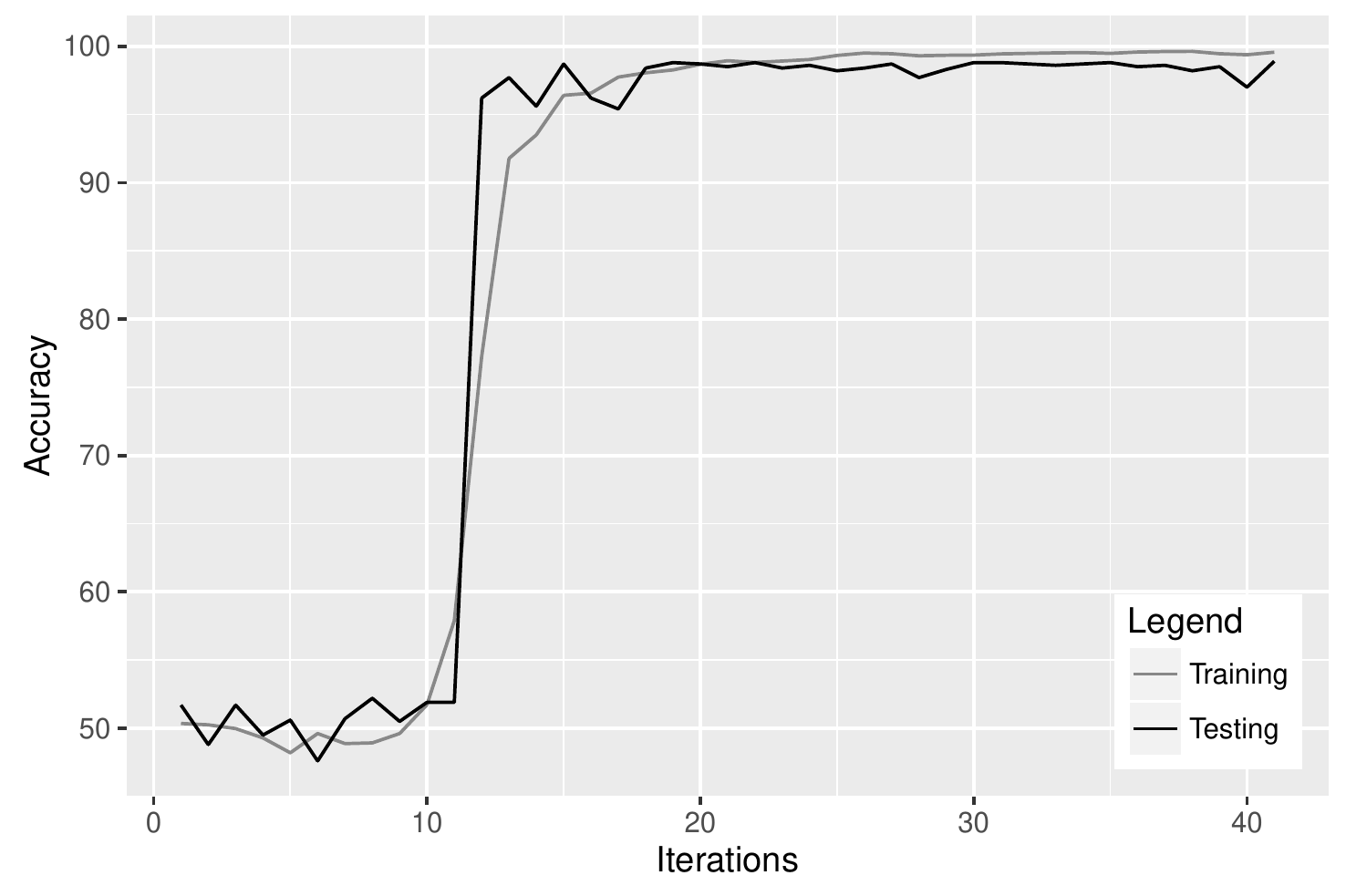}
}%
\hspace{4pt}%
\subfigure[][HUGO, $\alpha=0.1$~bpp]{%
\label{fig:accuracy_HUGO_0.1}%
\includegraphics[scale=0.39]{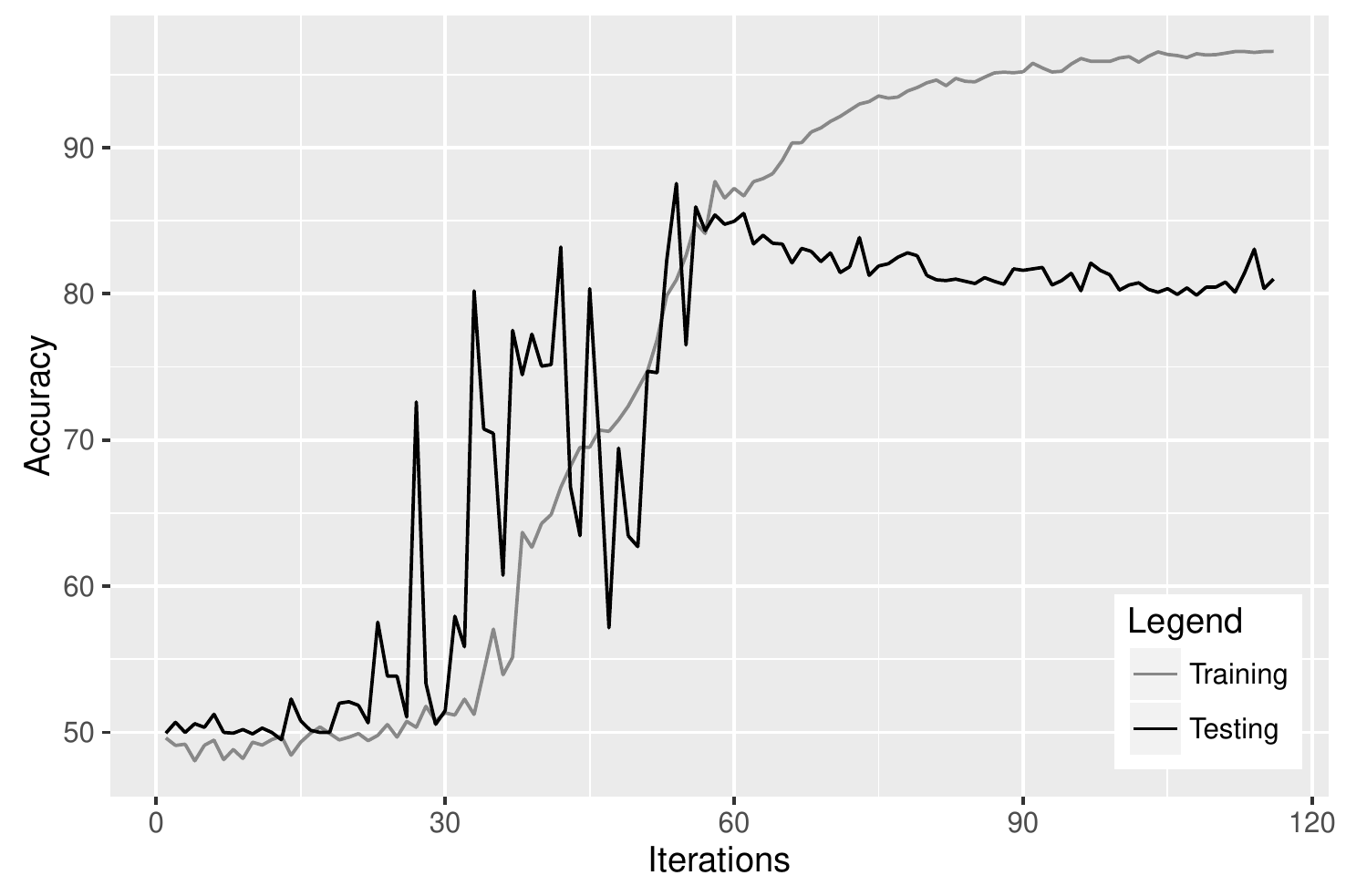}
}%
\\
\subfigure[][J-UNIWARD, $\alpha=0.4$~bpp]{%
\label{fig:accuracy_UNIWARD_0.4}%
\includegraphics[scale=0.39]{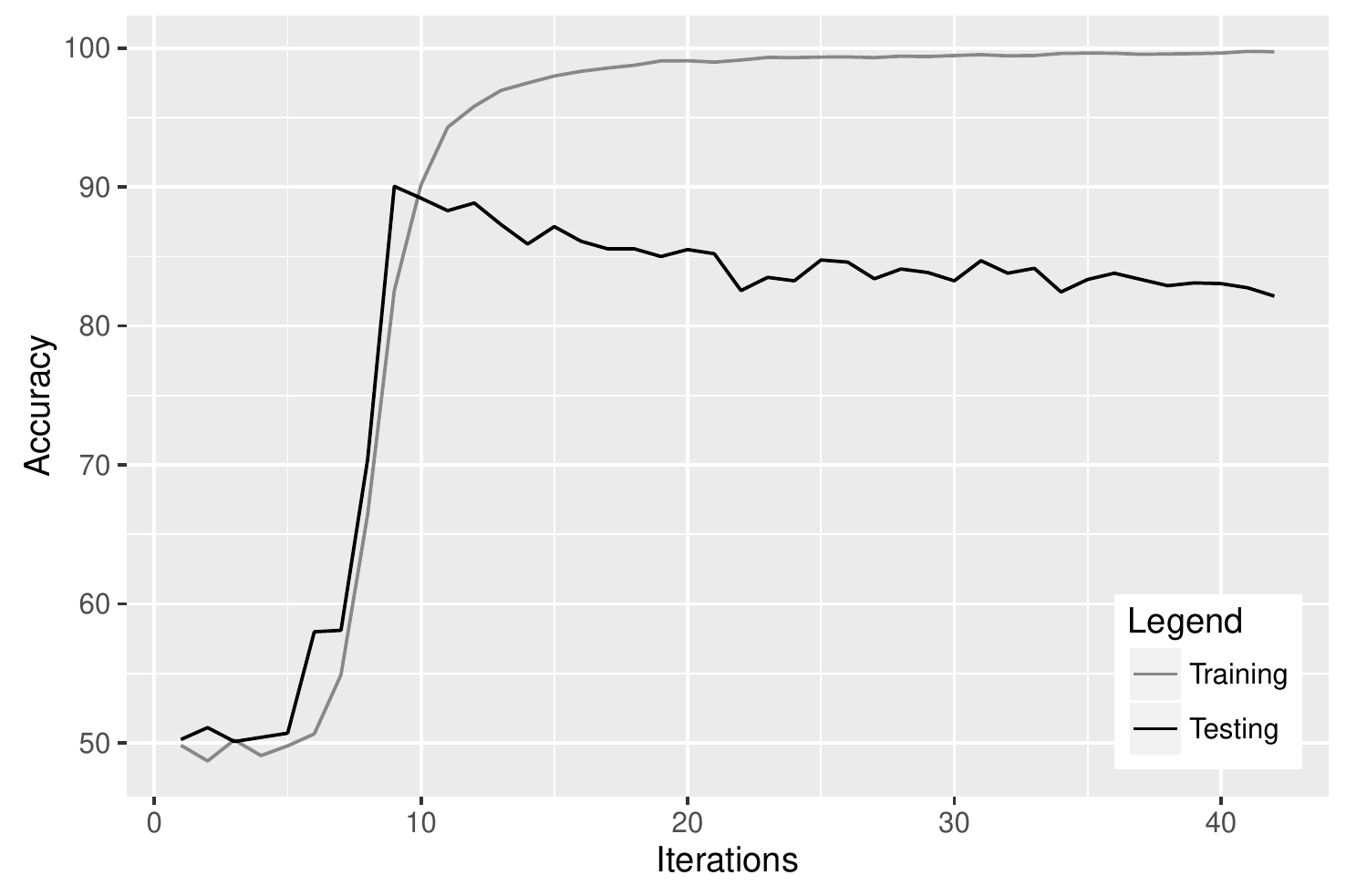}
}%
\hspace{4pt}%
\subfigure[][J-UNIWARD, $\alpha=0.1$~bpp]{%
\label{fig:accuracy_UNIWARD_0.1}%
\includegraphics[scale=0.39]{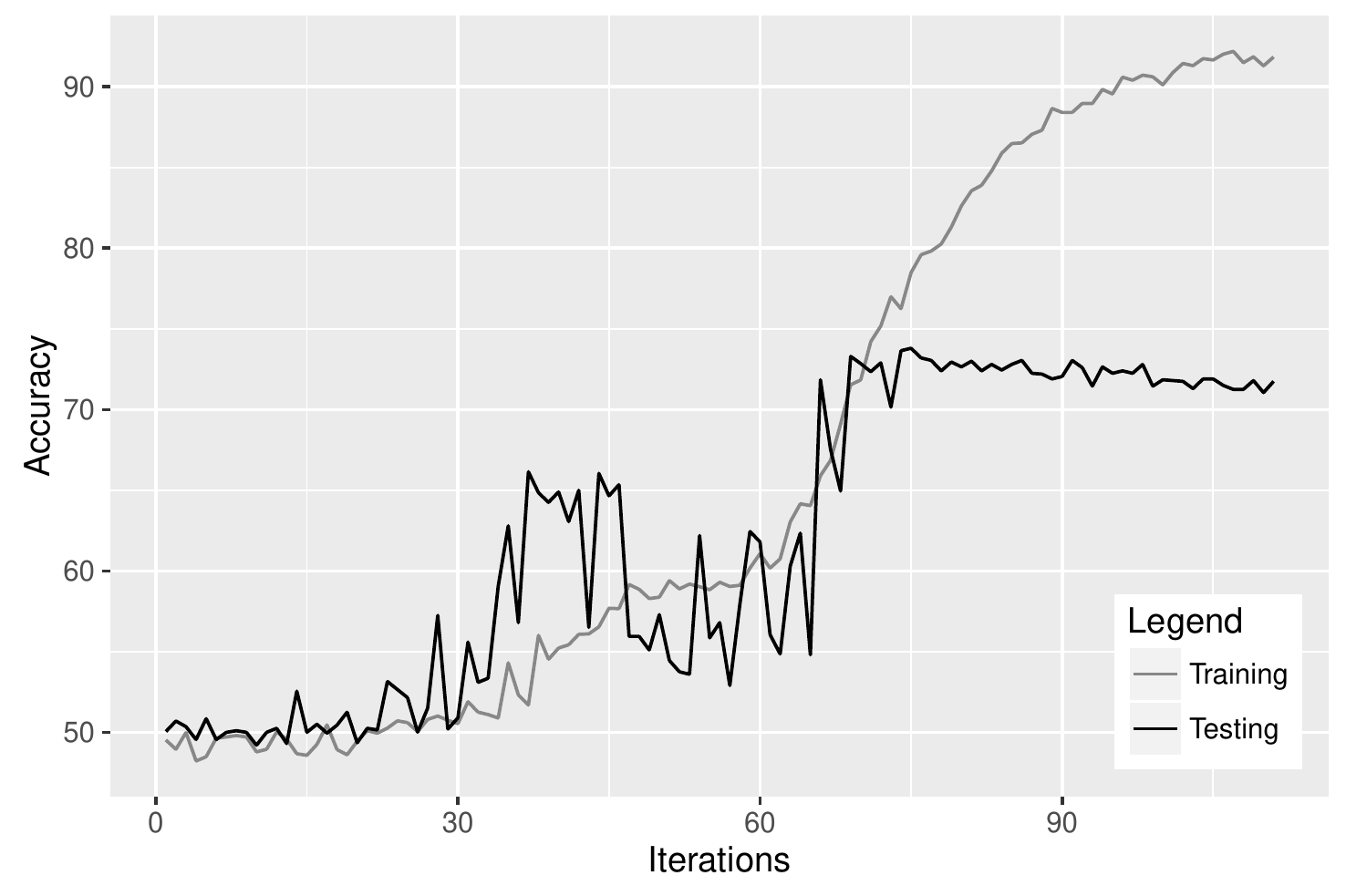}
}%
\caption[]{
Average detection accuracy as a function of the training iterations, and for both training and testing~datasets in case of WOW, HUGO, and J-UNIWARD steganographic schemes (with a payload $\alpha$).}
\end{figure}

\enlargethispage{0.25cm}

Figure~\ref{fig:accuracy_WOW_0.4} shows the evolution of the detection accuracy, on both training and test images, obtained for WOW with payload of~0.4~bpp. We can see that both curves are relatively similar: a fast convergence with a great improvement in a few iterations (epochs). For the lower payload value of 0.1~bpp, Figure~\ref{fig:accuracy_WOW_0.1} highlights a slower training convergence, as one would expect. We can also see that the testing accuracy exhibits a faster increase than the training one at the beginning, reaching after large oscillations a maximum value near iteration~50, before it slowly decreases to find a stabilized value a little higher than 80\%. Similar observations can be made for HUGO according to Figures~\ref{fig:accuracy_HUGO_0.4} and~\ref{fig:accuracy_HUGO_0.1}, which respectively show the curves obtained for HUGO with a payload of 0.4 and 0.1. Finally, in the case of J-UNIWARD we can note that even if Figures~\ref{fig:accuracy_UNIWARD_0.4} and~\ref{fig:accuracy_UNIWARD_0.1} show globally the same evolution for both accuracies than in the corresponding figures for WOW and HUGO, some slight differences appear. Firstly, the testing accuracies reach for J-UNIWARD a lower final value, with an accuracy above 70\% for a payload of 0.1, whereas it was above~80\% for the former steganographic algorithms. Secondly, on the one hand the training convergence is slightly faster for the largest payload, whereas it is slower for the low payload value. We think that two reasons can explain these observations: J-UNIWARD embeds the secret message in the frequency domain and not in the spatial one like WOW or HUGO, and also for J-UNIWARD the payload is defined as the percentage of modified non zero coefficients of the DCT values.

One might criticize the choice we made for the training process stopping criteria, since the testing curves in all figures clearly suggest that an early stopping is more suitable. This is especially true when the training of the networks is quite time consuming, as it is the case for the proposed CNN due to the huge number of parameters to optimize. In practice, one iteration takes approximately 40~minutes on a NVIDIA~K40~GPU, resulting in a training of more than 3~days for 110~iterations. A better stopping strategy for the training process would allow to greatly reduce this computation time by avoiding needless iterations. Using simultaneously several GPUs is also a solution to reduce the training time. However, optimizing the training process and the corresponding computation times was not our main concern in this study. It was rather to design a CNN that outperforms other steganalyzers regarding the detection accuracy.

\item Scenario~2

In this second scenario, we have chosen some of the previously trained networks and we applied them on a testing dataset built from 8,000 images taken from the Raise database. Each image plays the role of cover and stego after its processing by a steganographic algorithm (to embed a hidden message), thus we have a testing database consisting of 16,000~images. From the image content point of view, the Raise database is rather different from the BOSS one, therefore this scenario can be seen as an equivalent to the cover-source mismatch scenario in \cite{pibre2016deep}.

To start we consider the different final CNNs obtained in scenario~1 after training. Table~\ref{tab:01} reports the detection accuracy observed for these networks on the testing database. In this table, the two first columns present, respectively, the steganographic algorithm and the payload value used to generate stego images for the training and testing databases. The third column contains the iteration number during which the parameters of the CNN were obtained, and is followed by the last three columns giving the detection accuracy for the cover and stego images, and the total accuracy for the whole test set. On the one hand, we can see that the accuracy is very high for the testing stego images embedding information with a payload of~$0.4$, whatever the embbeding scheme and the payload value used for the training. In the case of the lower payload of~$0.1$, despite a decrease, since it is more difficult to track the information added by the steganographic algorithm, the accuracy remains acceptable. On the other hand for the cover images, the detection accuracies are always worst than for the stego ones, and particularly not so good when the training is done with the low payload value. We think that this scenario is, nevertheless, very interesting because stego images are in most of the cases detected, even if there are many false negatives.

%The second scenario consists in testing with 8,000 images issued from the Raise dataset. For these experiments, we have chosen some of the previously trained networks.

%In Table~\ref{tab:01}, experiments are reported in which the last network has been used for each experiment for the training. Column three shows the number of the network. In this table, we can see that the accuracy on the stego images is very high with a payload of 0.4. With a payload of 0.1, it is still very good. For the cover images, with a payload of 0.1, the accuracy is not so good. We think that this scenario is, however, very interesting because stego images are in most of the cases detected, even if there are many false negatives.

\begin{table}[t] %[htbp]
\centering
\begin{tabular}{|c|c|c|c|c|c|}
%\hline
%network       & testing      & number &  \multicolumn{3}{c|} {\% accuracy}  \\
%\cline{4-6}
% trained with  &              & network   & cover           & stego     &  total \\
%\hline
\hline
\multicolumn{2}{|c|}{Steganographic algorithm and payload} & Iteration &
  \multicolumn{3}{c|}{Detection accuracy} \\
\multicolumn{1}{|c}{Training} & \multicolumn{1}{c|}{Testing} & number & \multicolumn{1}{|c}{Cover} & \multicolumn{1}{c}{Stego} & \multicolumn{1}{c|}{Total}\\
\hline
WOW 0.4              &  WOW 0.4 & 42   & 91.16\% & 99.73\% & 95.44\% \\
\hline
WOW 0.1              &  WOW 0.1  & 108  & 59.39\% & 89.80\% & 74.59\% \\
\hline
WOW 0.1              &  WOW 0.4  & 108  & 59.39\% & 99.56\% & 79.47\% \\
\hline
HUGO 0.4             &  HUGO 0.4 & 41  & 94.66\% & 99.53\% & 97.09\% \\
\hline
HUGO 0.1             &  HUGO 0.1 & 116  & 60.50\% & 88.68\% & 74.59\% \\
\hline
HUGO 0.1             &  HUGO 0.4 & 116  & 60.50\% & 99.68\% & 80.09\% \\
\hline
J-UNIWARD 0.4       & J-UNIWARD 0.4 & 43 & 91.16\% & 99.73\% & 95.44\% \\
\hline
J-UNIWARD 0.1       & J-UNIWARD 0.1 & 111 & 52.19\% & 89.54\% & 70.86\% \\
\hline 
J-UNIWARD 0.1       & J-UNIWARD 0.4 & 111 & 52.19\% & 99.48\% & 75.81\% \\
\hline
\end{tabular}
%\caption{Results with the last network during the training}
\caption{Steganalysis results obtained from the final CNNs produced by the training process under scenario~1.}
\label{tab:01}
\end{table}

Now, if we replace each final convolution network by the best one (the optimal CNN from the testing accuracy point of view) find throughout the training, we obtain the results reported in Table~\ref{tab:02}. It can be noticed that the best network is usually identified well before the end of the training, as already highlighted by the figures from scenario~1. The exception is an embedding using HUGO with a payload of~0.4. A look at the total detection accuracy confirms that the retained CNN are almost all better than in Table~\ref{tab:01}: the training with J-UNIWARD~0.4 is the only odd case. We think that this case is an outlier resulting from the test data, because we can see on Figure~\ref{fig:accuracy_UNIWARD_0.4} that the training accuracy is not very good when the testing one is optimal. The improvement of the total accuracy mainly comes from the highest accuracies observed for the cover images, whereas conversely, they are slightly lower for the stego images.

In both tables, we can see that a CNN trained with a small payload is able to detect images embedding information with a higher payload. Obviously, the lower the payload is, the more difficult the detection task is. To sum-up, for a payload of 0.4~bpp the proposed CNN can detect stego images with an accuracy higher than 98\%, whatever the steganographic algorithm chosen among WOW, HUGO, and J-UNIWARD, while it falls at most to 73.30\% for the payload of~0.1. In comparison with the experiments presented in \cite{pibre2016deep}, focused on the steganographic algorithm S-UNIWARD at 0.4~bpp, our quite different CNN is very competitive and outperforms conventional steganalyzers using a combination of a Rich Model and Ensemble Classifier for the same embedding key scenario.

%In Table~\ref{tab:02}, experiments are reported in which the network with the best training is used for the validation with images from the Raise dataset. In Table~\ref{tab:01}, experiments are reported in which the last network has been used for each experiment for the training. It can be noticed that for HUGO 0.4, the best network is the last one. So the result is the same than in the previous table. In this case, results are better than in the previous table. However, the probability of errors for stego images is lower. Consequently, the probability of error for cover images is larger.

%In both tables, we can see that a network trained with a small payload is able to detect images with a higher payload.

\begin{table}[t] %[htbp]
\centering
\begin{tabular}{|c|c|c|c|c|c|}
\hline
\multicolumn{2}{|c|}{Steganographic algorithm and payload} & Iteration &
  \multicolumn{3}{c|}{Detection accuracy} \\
\multicolumn{1}{|c}{Training} & \multicolumn{1}{c|}{Testing} & number & \multicolumn{1}{|c}{Cover} & \multicolumn{1}{c}{Stego} & \multicolumn{1}{c|}{Total}\\
\hline
WOW 0.4              &  WOW 0.4 &  34  & 94.96\% & 99.59\% & 97.28\% \\
\hline
WOW 0.1              &  WOW 0.1  & 52  & 74.46\% & 81.03\% & 77.74\% \\
\hline
WOW 0.1              &  WOW 0.4  & 52  & 74.46\% & 99.01\% & 86.73\% \\
\hline
HUGO 0.4             &  HUGO 0.4 & 41  & 94.66\% & 99.53\% & 97.09\% \\ %%meilleure valeur
\hline
HUGO 0.1             &  HUGO 0.1 & 54  & 79.90\% & 73.30\% & 76.60\% \\
\hline
HUGO 0.1             &  HUGO 0.4 & 54  & 79.90\% & 99.28\% & 89.59\% \\
\hline
J-UNIWARD 0.4       & J-UNIWARD 0.4 & 9  & 85.60\% & 98.46\% & 92.03\% \\    %%meilleur résultat en test sur boss mais pas ici sur raise => à commenter
\hline
J-UNIWARD 0.1       & J-UNIWARD 0.1 &  75 & 62.91\% & 83.91\% & 73.41\% \\ 
\hline 
J-UNIWARD 0.1       & J-UNIWARD 0.4 &  75 & 62.91\% & 98.52\% & 80.72\% \\ 
\hline
\end{tabular}
%\caption{Results with the best network during the training}
\caption{Steganalysis results obtained from the best CNNs produced by the training process under scenario~1.}
\label{tab:02}
\end{table}

\end{itemize}

%The training of networks is quite time consuming. One iteration take approximately 40 minutes. So for more than 110 iterations, it takes more than 3 days. Using simultaneously several GPUs is a solution to reduce the training time. In the github code, we give the pre-trained networks in order to be able to reproduce these experiments.

%\begin{color}{red}
%C'est dit deux fois dans l'article
% \end{color}

\section{Discussion} \label{sec:discuss}

Obviously, the widespread reuse of the same embedding key does not only undermine the security of the steganographic algorithms, but it also makes easier for a steganalysis algorithm to detect images embedding a secret message. Moreover, since the stego key is used to locate the embedding positions on a cover image, a single key greatly restricts the choice of the pixels which will be modified to hide the message. Steganalysis tools that exploit spatial correlation to detect stego images, like convolutional networks, are thus favored and that explains the obtained results. And besides, when the embedding is done considering different keys the detection accuracy of our CNN drops dramatically and is clearly outperformed by the conventional steganalysis scheme based on Rich Models and Ensemble Classifier. Despite this major drawback resulting from the misuse of embedding simulators, several remarks can be made on this work.

First of all, this study that extends the findings of the previous works on steganalysis using deep~learning, strengthens the idea that this branch of machine learning, and more particularly the convolutional neural network architecture, is a key research direction for the design of steganalyzers. However, as the proposed CNN is quite different from the ones resulting from earlier works \cite{qian2015deep,pibre2016deep}, our work actually questions some assertions made in \cite{pibre2016deep} on the design of convolutional neural networks for steganalysis. More precisely, we have shown that the preliminary upstream filtering with $F^0$ of the input image before the CNN is not mandatory, but can be replaced by a global filtering done in the first convolution layer, and that a huge number of features is not needed to accurately detect stego images. 

We think that the success of our proposal for the same key scenario mainly comes from the use of large convolution filters. Indeed, such large filters build small long-range correlation patterns and thus allow to obtain a reduced set of well-discriminating features (256~features for $512 \times 512$~pixels images). The relevance of the proposed convolutional part is further supported by the fact that the classification task can be fulfilled with a high accuracy by a fully connected part reduced to its simplest form consisting of the output neurons. Different steganographic tools, working in spatial or frequency domains, can be detected with almost no error for a payload of 0.4~bpp, but the detection ability is less satisfactory for the lower payload value of 0.1~bpp. As expected, the lower the payload value is, the more difficult the detection of stego images will be due to the reduced number of modified cover pixels. As an illustration, Table~\ref{tab:payloads} summarizes the average number of pixels that have been modified, with respect to the selected payload, on the 10,000 BOSS $512\times 512$ images. For instance, when the payload is set to $\alpha=0.1$ with HUGO, only 2,425 pixels are modified on average in each image whereas $0.1 \times 512\times 512$ is 26,214. The difference is mainly due the use of the Syndrome Treillis Code~\cite{DBLP:conf/mediaforensics/FillerJF10}.

\begin{table}[t]
    \centering
    \begin{tabular}{|c|r|r|r|}
        \hline
         Payload & \multicolumn{3}{|c|}{Steganographic algorithm} \\
         value & \multicolumn{1}{|c}{WOW} & \multicolumn{1}{c}{HUGO} & \multicolumn{1}{c|}{J-UNIWARD} \\
        \hline
         0.1     &  4,714    &  4,872    &  727 \\
        \hline
         0.4     & 23,575    &  23,557    & 3,591         \\
        \hline
    \end{tabular}
    %\caption{Average number of modified bits w.r.t payloads.}
    \caption{Average number of modified pixels w.r.t payloads.}
    \label{tab:payloads}
\end{table}

Some preliminary experiments made in the case of training-testing stego~images mismatch show that some networks are able detect multiple steganographic algorithms. For example, we have observed that a CNN trained with WOW is able to detect images embedding data using HUGO. Therefore, it would be interesting to investigate this point with other steganographic tools, in particular tools working in the same domain.

An early stopping approach would allow to avoid overfitting and greatly reduce the number of wasteful training epochs, and thus the computation time spent to train a network. But the design of such a stopping strategy must be done carefully, in order to prevent a premature stopping. The benefit of a powerful computing platform is also important, even if after the training of a network with a good GPU, like a~K40 in our case, a less capable GPU can be used for the testing. As an illustration, a~K610 with only 192~cores and 1~GB of memory can run all the tests presented in this article.

%Some networks are able to detect multiple steganographic tools. For example, we have seen that a network trained with WOW is able to detect  images steganographied with HUGO and reciprocally. We think it would be interesting to investigate this with other steganographic tools.

%After the training of a network with a good GPU, a K40 in our case, a small GPU can use it for the testing. For example, a K610 with only 192 cores and 1GBytes of memory can run all the tests of this paper.

\section{Conclusion and future work} \label{sec:conclusion}

Detecting steganographic content in images is a difficult task, which, up to now, was mainly addressed by the use of a Rich Model with an Ensemble Classifier. Recently, the increasing attention gained by deep learning approaches, due to breakthrough results on various challenging classification tasks, has raised the question of whether such an approach is relevant for the design of steganalyzers. Beginning with the pioneer work of Qian {\em et al.}, later continued by Pibre {\em et al.}, the investigation of this question has shown that convolutional neural networks are promising for steganalysis. In this article, we further strengthen this research direction towards CNN-based steganalysis by the design and the very positive evaluation of such deep neural networks. However, the conclusions drawn by this work and the previous earlier ones are restricted by the use of a single embedding key. A major security flaw that may occur when the user has a bad understanding of the underlying embedding process in a steganographic algorithm, or due to improper use of a steganography software tool.

The proposed CNN has a quite different shape compared to the ones resulting from the earlier works, and it is able to provide high detection accuracies for several steganographics tools when the same stego key is reused during the embedding process. The convolutional part of our proposal starts by a global filtering, using a single filter, followed by a second convolutional layer that produces a reduced set of high-level features (256~features for $512 \times 512$~pixels input images) thanks to the use of large filters. The information encoded by the final vector of features is so discriminating that the classifier part can be reduced to only two output neurons. We finally evaluated the detection ability of the CNN against two spatial domain steganographic schemes and a frequency domain one. The obtained results are very encouraging, and they outperform all the previous deep learning proposals for steganalysis. More precisely, we designed a perfect steganalyzer for embedding payloads of 0.4 bit per pixel, and for all the steganographic tools investigated in this article (working either in spatial or in frequency domains). Rather interesting results have been obtained too, albeit to a lesser extent, for a payload value of 0.1 bpp. Finally, a first step in the design of a universal detector has been achieved here, as we are able to detect HUGO based hidden messages even when a WOW steganographier has been used during the training stage.

In future work, our intention is to enlarge the set of steganographiers considered during both the training and the testing stages, and to study other frequency domains for embedding, like the wavelet one. These tools will be mixed in various scenarios, in order to make our steganalizer as insensitive as possible to the embedding process (universal blind detection). To speed up the process, we will also incorporate an early stopping control element during the learning stage. Finally, theoretical investigations will be considered, in order to have a better understanding of the reasons why such a CNN works so well. To do so, we will, in particular, investigate the shape of the filters found by the convolutional neural network, and try to relate their coordinates to the chosen steganographiers.

\section*{Acknowledgments}

This article  is   partially  funded  by  the  Labex   ACTION  program  (ANR-11-LABX-01-01 contract) and the Franche-Comt\'e regional council.  We would like to thank NVIDIA for hardware donation under CUDA Research Center~2014 and the Mésocentre de calcul de Franche-Comté for the use of the GPUs.

\bibliographystyle{unsrt}
\bibliography{references}
\end{document}